# A strategy to identify breakdown location in MITICA test facility: results of high voltage test campaign

Loris Zanotto[1], Marco Boldrin[1], Giuseppe Chitarin[1], Mattia Dan[1], Tommaso Patton[1], Francesco Santoro[1], Vanni Toigo[1], Hiroyuki Tobari[2], Atsushi Kojima[2], Hans Decamps[3]

[1]*Consorzio RFX (CNR, ENEA, INFN, Università di Padova, Acciaierie Venete SpA), Padova, Italy*
[2] *National Institute of Quantum and Radiological Science and Technology, Mukouyama 801-1, Naka 319-0193, Japan*
[3] *ITER Organization, Saint Paul lez Durance, France*

*Abstract –* The Acceleration Grid Power Supply of the MITICA test facility in Padova (Italy) is currently under commissioning. The power conversion system, the DC generator, and the High Voltage equipment have been individually commissioned, whereas the integration tests are ongoing. It is a challenging process due to the unconventional application, to the variety of different electrical technologies involved and to the complexity of the interfaces.
During the integrated tests of the power supplies the achievement of 700kV stable operation has been demonstrated for the first time in a Neutral Beam Injector, but an unexpected event occurred, most likely a breakdown in the HV part, which resulted in a fault of the DC generator. A subsequent test using an auxiliary power supply was performed to check the voltage withstanding capability of the HV plant, but another breakdown occurred at around 1MV.
This paper describes the activity performed to identify the location of the breakdowns affecting the integrated tests. A test campaign has been devised with increased diagnostic capabilities and specific strategy conceived to trigger intentional breakdowns in specific locations and collect measurement patterns for different cases. The results of the campaign will be presented and the current understanding of the issue will be described, with a view on future tests and further improvements of diagnostics.

## 1. Introduction

The full scale 1MV Neutral Beam Injector (NBI) MITICA (Megavolt ITER Injector & Concept Advancement) test facility [1] includes a very complex and special power supply system, comprising a High Voltage DC plant with some parts insulated in air, some other in SF6 gas and in vacuum. The system is designed to test the injector concept and technology in advance of ITER operation and reach the required performance, i.e. extracting, accelerating and neutralizing Deuterium and Hydrogen ions up to -1MV (-870kV for Hydrogen) and 40A (46A for Hydrogen). An overview of the power supply system is shown in Fig. 1. The main power supply is the Acceleration Grid Power Supply (AGPS, see 1 in Fig. 1) [2], feeding around 56 MW at -1MV dc to the acceleration grids: five stages, each composed of an oil insulated step-up transformer and a diode rectifier and rated to produce 200kV dc, are connected in series at the output and to a RC filter unit (DCF, see 2 in Fig, 1). The diode bridges and the DCF are installed inside $SF_6$ pressurized tanks. The step-up transformers and the diode bridges are indicated as AGPS DC Generator (AGPS-DCG). A large, -1MV dc air insulated Faraday cage, called High Voltage Deck 1 [3] (HVD1, see 3), hosts the power supplies of the Ion Source [4], called ISEPS (Ion Source and Extraction Power Supplies), which are sitting at -1MV potential, devoted to provide power for the production and extraction of the ions. The ISEPS are fed by an Insulating Transformer (see 4). A 100m long Transmission Line (TL, see 6), divided into three sections TL1, TL2 and TL3, insulated in SF6 at 0.6MPa abs, connects ISEPS and AGPS to the beam source and the accelerator inside the MITICA vessel (see 8) via an $SF_6$-to-vacuum bushing (see 7). ISEPS conductors are routed to the TL through an air-to-$SF_6$ bushing placed below the HVD1 (see 5, this bushing is called High Voltage Bushing Assembly, HVBA)

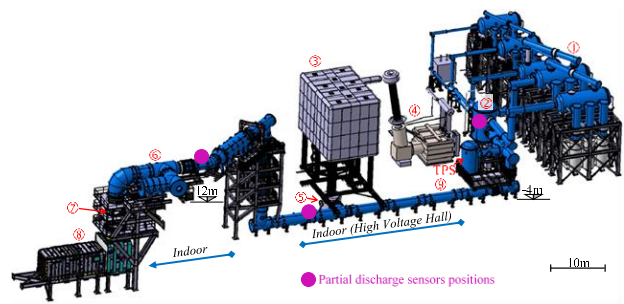

Fig. 1 – Overview of MITICA Power Supply system.

A so-called Short Circuiting Device (SCD) is installed in place of the beam source to perform commissioning and integration tests of the power supplies. The SCD, insulated in $SF_6$ at 0.13 MPa abs, is equipped with five movable spark gaps which can be controlled remotely, to simulate grids breakdown (BD); five fixed parallel spark gaps are installed to limit the voltage to admissible value (-1 MV dc). The MITICA power supply system is extremely complex as it involves many different technologies at the boundary of technical feasibility, at unprecedented power and voltage level, with many interfaces (mechanical, electrical, cooling, signal). Such

*author's email: loris.zanotto@igi.cnr.it*

complexity is one of the main reasons why the MITICA test facility was built, i.e. to test and troubleshoot the system in advance of ITER, so that criticalities can be found in time and necessary provisions designed and implemented before the construction in ITER. The commissioning of the system has followed a step-by-step approach, due to the complexity of the interfaces and to the different stakeholders and suppliers involved in the procurement, which was shared mainly between the European and Japanese domestic agencies. Acceptance tests were carried out and successfully passed individually for each item. In particular, for the HV equipment, the voltage holding tests have been performed between 2018 and 2019 using an auxiliary power supply called Testing Power Supply (TPS) rated for 1.3MV, 10mA. Details of these tests can be found in [5,6]. In 2021, the integration tests of the AGPS were started with a campaign aimed at integrating the AGPS Conversion System (AGPS-CS) [7] with the High Voltage equipment. For these tests the AGPS was connected to a special Dummy Load (DL) installed indoor nearby the HVD1. These integration tests demonstrated the ability of the system to generate and sustain 700kV for 1000s successfully for the first time in a NBI, which is very important in view of ITER. However, an unexpected event occurred while operating the system at around 800kV, most likely a breakdown in the HV plant, which resulted in a fault of the DC generator, precisely in the diode bridge of stage 5. A subsequent test using the TPS was performed to check the voltage withstanding capability of the HV plant, but another breakdown occurred at around 1MV. In both cases, an initial breakdown event generated a voltage surge which propagated along the Transmission Line, causing electrical stresses and damages on the equipment. Analyses and inspections showed that, although the origin of the voltage surge was likely to be a breakdown in the HV equipment (either in air or in gas), its precise location was unclear and not fully identified. After a description of the events occurred in 2021 and an overview of the investigation on the dynamic of the voltage surge leading to the failure of the components, this paper presents the strategy, the preparation activities and the results of a test campaign aimed at discriminating the position of the breakdown event which originated the voltage surge. The results will be discussed in the conclusions with a view on future tests and enhancement of the diagnostic capabilities in the High Voltage plant.

## 2. Integration tests

Fig. 2 shows a simplified diagram of the MITICA HV circuit during the integration tests of the AGPS, started in early 2021. The AGPS-CS was connected to the AGPS-DCG. The power was delivered to each stage by means of inverters, which could be controlled to modulate the dc voltage at the output of the AGPS-DCG. The voltage was applied to the resistive DL, which was connected between the HVD1 and the external TL conductor. The SCD, with movable contacts in the open position, was connected at the end of the TL in place of the ion source.

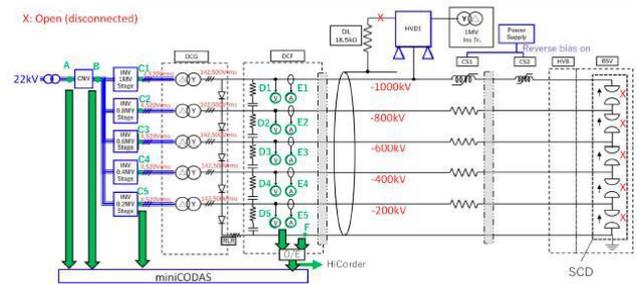

Fig. 2 – Simplified scheme of the MITICA HV circuit during integrated tests (no-load configuration).

After testing each individual stage at low voltage (up to 50kV per stage) to check the functionality, all five stages were connected together in series and the voltage was increased step by step, with increasing pulse duration, initially with DL disconnected (no-load operation). A typical pulse produced during such tests is reported in Fig. 3, where the voltage of each conductor with respect to ground is shown. The transient at the end of the pulse starts at inverter turn-off and represents the discharge of the circuit capacitance mainly to through the cooling water resistance.

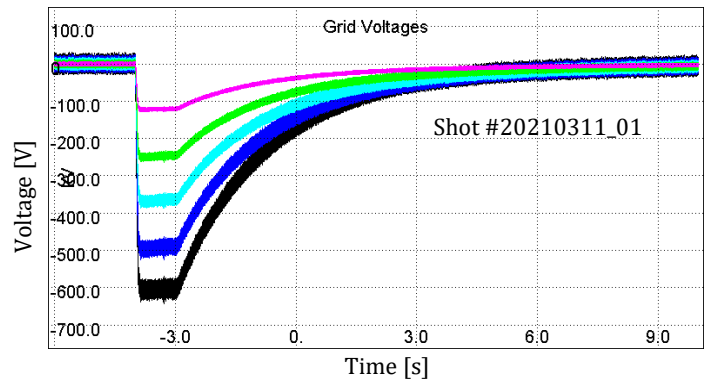

Fig. 3 – Voltage of the grids during pulse #20210311_01. Different colors represent the voltage of each grid with respect to ground, from -200kV (purple) to -1000kV (back) grid.

After some days of successful operation, where the capability of the system to sustain voltage as high as 700kV in stable operation for at least 1000s has been demonstrated, an event occurred during a pulse at 800kV, see Fig. 4. The event was experienced as a strong noise captured by the microphones placed in the High Voltage Hall (the building where the HVD1 and DL are installed). Looking at the grid voltages, which suddenly fell to zero, it was clear that a breakdown was triggered somewhere in the plant.

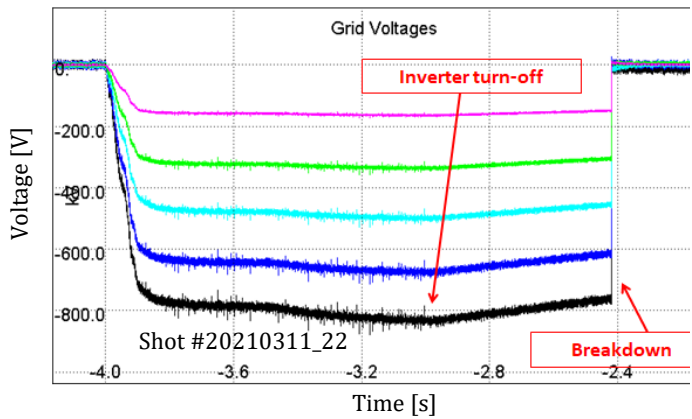

Fig. 4 – Voltage of the grids in pulse #20210311_22. Same color code as Fig. 3. A breakdown occurred at about -2.4s.

The following inspection of the plant did not show sign of BD neither in $SF_6$ insulated components nor in air, but a broken diode bridge arm in one DCG stage was detected. The analysis based on circuit simulations taking into account stray parameters of the DCG gave a convincing explanation of why the diode bridge was damaged, highlighting that the voltage surge caused by a BD in the plant could damage the first stage of the DCG due to excessive overvoltage. However, the position of the BD was not found in the inspection. Being the AGPS out of service due to the fault in the diode bridge, it was decided to check again the voltage withstanding capability of the HV equipment by using the TPS, connected in place of the AGPS-DCG. The test consisted in increasing the voltage in steps up to the nominal value and check that the equipment was able to sustain the voltage, confirming the results of the acceptance tests. However, immediately after reaching -1MV dc, an arc in air was triggered at the top of the -1MV Insulating Transformer bushing, precisely between the guard rings and the AC line feeder terminals of the secondary side of the transformer, as indicated in Fig. 5. Thanks again to a analyses based on circuit simulations [8,9] which considered the contribution of transformer and bushing stray parameters, it turned out that a large transient overvoltage (more than 1MV), able to trigger an arc, can appear between the guard rings and the ac line terminals in case of BD on other parts of the system.

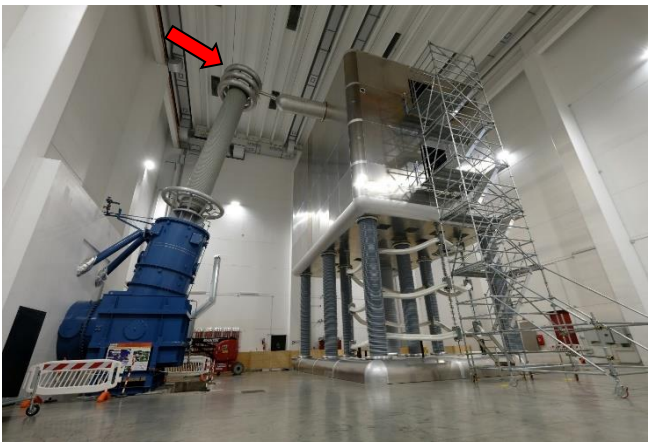

Fig. 5 – Picture of the HVH showing the position of the arc in air triggered during TPS test.

On the basis of these simulations, the most likely explanation appears to be that the arc was actually triggered by a first BD somewhere in the HV plant, which propagated an overvoltage at the top of the transformer bushing. Once triggered, the arc in air was sustained by the ac voltage until the opening of the protection breaker. Unfortunately, not all plant components could be inspected and none of the accessible components of the plant did show clear marks of BD, neither in air nor in $SF_6$. Nevertheless, the integration tests were very useful to highlight that a BD in the plant is able to cause unexpected overvoltage on some components: this demonstrates the clear advantage of testing the HV equipment in MITICA in advance of the ITER installation. The location of the BD must be identified in order to get a full understanding of the phenomena and take the necessary mitigation actions. This is the reason why a strategy based on a new test campaign with increased diagnostic capabilities has been conceived as described in the next section. This strategy will be tested in MITICA but will give important information for the ITER NBI also.

### 3. A strategy to identify the breakdown location

#### A. Diagnostic improvements

The first step towards the identification of the BD location was the improvement of the existing set of plant measurements and diagnostics. The idea was to add electric measurements, special cameras, microphones, partial discharge and corona sensors in specific parts of the MITICA plant, with the aim of collecting useful information possibly giving a suggestion on the BD location and helping in the understanding of the dynamics of the phenomena. Fig. 6 shows a scheme of the MITICA plant with an overview of the diagnostic systems and electrical measurements installed to this purpose. As a first step, it was particularly important to understand whether the weak insulation points were in the air or $SF_6$ parts, therefore the cameras and sensors already installed in HVH were complemented by a set of partial discharge sensors, field probes, acoustic sensors and new cameras suitable for DC applications, identified by CESI [10]. The idea was to add electric measurements, special cameras, microphones, partial discharge and corona sensors in specific parts of the MITICA plant, with the aim of collecting useful information, The system and its assessment, calibration and installation are described in [6]. On top of the sensors installed to monitor the air-insulated part, a large number of electric measurements were installed on various parts of the plant, i.e. rogowski coils, current transformers, voltage probes, to capture voltage and current transients related to a BD. Finally, two visible cameras were mounted on vessel ports to look at the SCD contacts: one "fast" high frequency camera with acquisition rate up to 1MHz (Phantom V2012 [11]) but with limited field of view and one "slow" camera with lower acquisition rate but broader viewing angle (Basler camera [12]).

Figs. 7a and 7b shows the installed fast camera and an example of viewing angle when the camera is set at 130 kHz sample rate and 480x640 pixel. The camera was actually set to the highest available sample rate (1MHz) in order to be able to capture the arc with a time resolution of 1µs; of course in this case the viewing angle is limited and can be adjusted to look at just one stage of the SCD.

Table I lists the sensors and measurements installed to improve the MITICA plant diagnostic capability.

### B. Design of a specific test campaign

The improvement of the diagnostic system enabled a strategy based on a new test campaign specifically devised for discriminating the BD location, after disconnecting the damaged parts. The test campaign has been conceived in different stages, first using an external high voltage power supply and then the TPS at increasing voltage up to 1MV. The idea was to trigger intentional short-circuits with a spark-gap in different positions and with the SCD in vessel in order to obtain benchmark waveforms, i.e. to identify at low voltage the pattern of measurements associated to different breakdown locations, so that the identification of the breakdown location in case of a non-intentional breakdown can be facilitated by comparing the measurements with previously benchmarked cases.

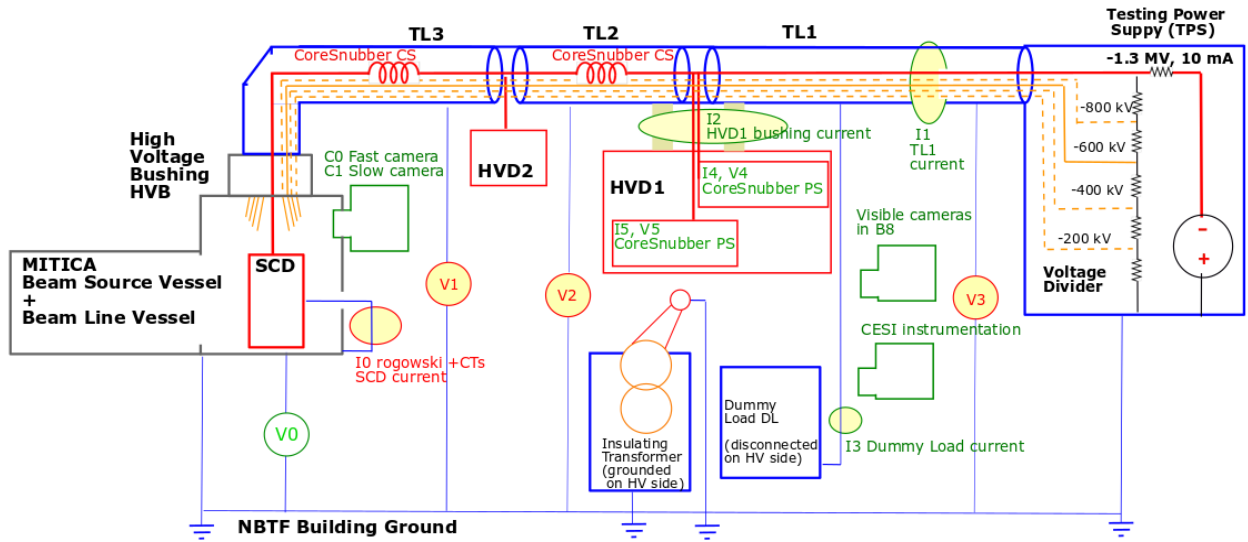

Fig. 6 – Scheme MITICA HV equipment with improved diagnostic system for breakdown localization.

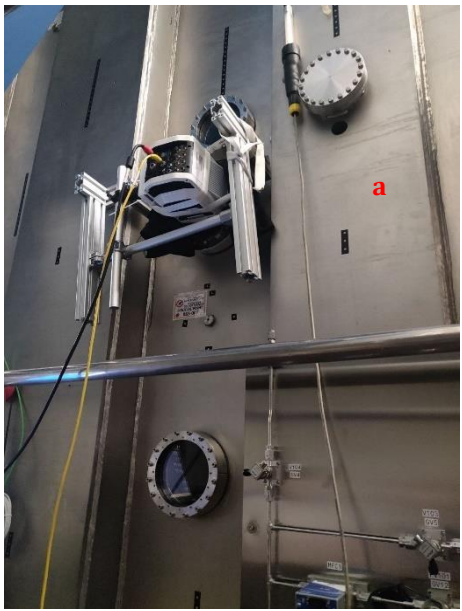
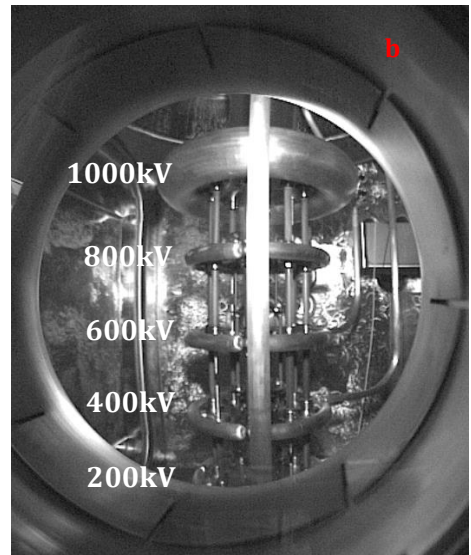

Fig. 7a,b – Picture of fast camera installation (a), and viewing angle for 130kHz sample rate and 480x640 pixel resolution (b).

Table I – List of electrical measurements and sensors installed in MITICA plant to detect BD location (HVH sensors [6] are not listed).

Table I – New measurements and sensors installed in the MITICA power supply plant.

| ID | Description | Type |
|---|---|---|
| V0 | Vessel to ground voltage | P6015 voltage probe |
| I0 | SCD current: current in connection between last stage and vessel. | 2 redundant CTs |
| C0 | Fast Camera | Phantom V2012 |
| C1 | Slow Camera | Basler |
| I1 | TL1 common mode current | Rogowski coil around the outer conductor |
| V1 | TL3 tank to ground voltage | Voltage divider |
| I2 | HVD1 common mode bushing current | Rogowski coil around the HVBA |
| V2 | TL2 tank to ground voltage | Voltage divider |
| I3 | Dummy Load current | Rogowski coil around connection between DL and TL tank |
| V3 | TL1 tank to ground voltage | Voltage divider |
| I4,V4 I5,V5 | Core snubber bias voltage to ground and currents | P6015 voltage probe and CTs |

To reach the objective, the campaign has been organized in three phases:

- Phase 0: Tests with external generator with TL in air, external spark gap and SCD, up to 50kV;
- Phase 1: Tests with external generator with TL in SF6 and SCD, up to 150 kV;
- Phase 2: Insulating tests with the TPS up to -1MV and simulation of grid breakdown by using the SCD. In this phase the instrumentation in the HVH has been moved around to detect signs of corona or partial discharge activity.

The aim of Phase 0 is to collect typical waveforms from current and voltage sensors during a breakdown in different locations and to check the electrical stress in different positions, including auxiliary systems. An external generator has been used to apply voltage up to 50kV. An external spark-gap has been connected in different positions, with the TL in air to facilitate access to the equipment. The identified positions for the intentional breakdowns were (see Fig. 8):

P1) HVD1 to ground, using a spark gap;
P2) HVD1 to DL, using a spark gap;
P3) TL2 to TL tank, using a spark gap;
P4) TL1 to tank, using a spark gap;
P5) SCD to vessel, using the SCD.

Phase 1 is aimed at checking the operation of plant up to a safe level of voltage (150kV) supplying the circuit with the TPS and using the SCD to trigger an intentional breakdown for each voltage step; current and voltage sensors has been checked to collect additional benchmark waveforms. In order to perform these tests, the TL must be insulated in SF6.

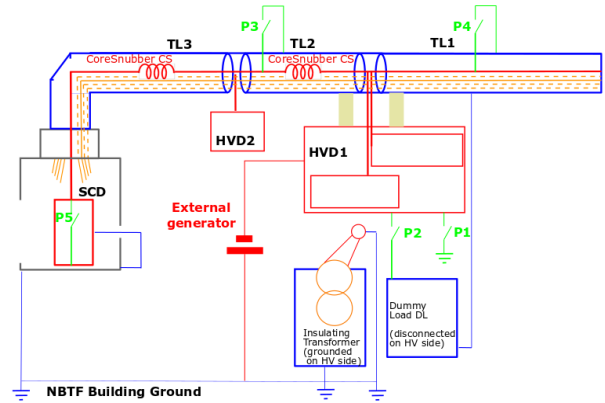

Fig. 8 – Positions of spark-gap short circuit for phase 0 tests with external generator.

Finally, Phase 2 consists of increasing the voltage up to -1MV, in order to find the position of weak points of the insulation and, if possible, identify the position of previous faults on the basis of a comparison of the experimental waveforms with the benchmark waveforms obtained in Phase 0 and Phase 1. In addition, the sensors identified by CESI and installed in HVH [6] have been moved around to detect possible weak positions of the air insulation, in search of precursor of discharge such as corona.

## 4. Test results

### A. Phase 0 & Phase 1

The external voltage generator has been used to apply voltage up to 50kV from 0, in steps of 10kV. A short circuit has been triggered in different plant positions; since the calibration at very low voltage is difficult, the spark-gap has been used at 50kV only, while for position P5 the SCD has been used for lower voltage levels also. Concerning the SCD, it was noticed that, especially at low voltage, the jitter in the closing time of the contacts in the five stages affects the arc generation, so that the short-circuit is actually happening not at the same time on different stages. In general, the tests gave a pattern of measurements which was distinctive for different breakdown location. As an example of waveforms obtained during such tests, Fig. 9 shows I2 current (rogowski coil measuring HVD1 common mode bushing current) for breakdowns in positions P2 and P3 at 50kV. Being the measurements very noisy, the signals have been filtered with a moving-average window filter based on the sensor rated bandwidth. The repeatability of the results has been confirmed by repeating the same short-circuit test several times.

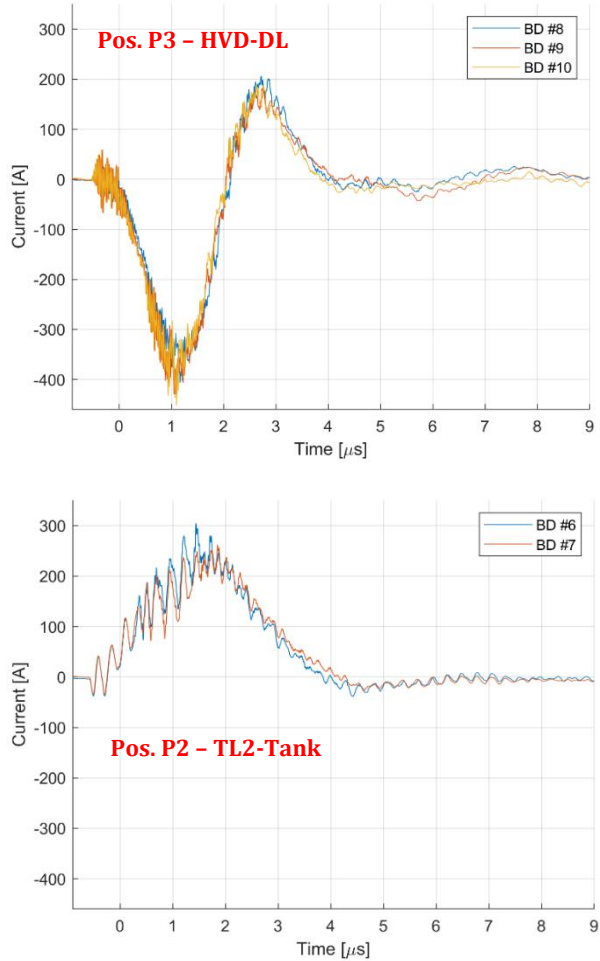

Fig. 9 – Current measured by sensor I2 for two different positions of the BD at 50kV; same pulse is repeated several times.

The qualitative features of different measurements have been collected in a table in order to find out whether it was possible to discriminate the BD position just looking at the qualitative patterns.

Table II reports the results in terms of sign of the first peak for each measurement for different position of the provoked BD. The sign of first peak is a solid indicator because it depends on the BD position, while the following oscillations depend on dissipative effects, not easy to be identified. Only a subset of the most meaningful measurements is shown in the table, but the results are available for all the sensors. Where the measured current/voltage is negligible or noisy, it is reported as ~0 or "N". For instance, for a BD between HVD1 and ground (position P1) we have found that the first peak of filtered I1 all I2 is negative while all other signals have negligible values; instead in case of BD between SCD and ground (position P5), I1 is negligible, first peak of I2 is positive, I0 and V0 shows large noise, while all other signals are negligible. Thus the sign of the first peak of the waveform is considered a reliable indicator of a BD position: in fact, it is little dependent on dissipative effects, which are usually difficult to be precisely quantified and affect the evolution of the transient in a longer timescale; the sign of the first peak can be thus modeled with relatively high confidence to confirm the experimental evidence.

Table II – Results of phase 0 tests in terms of measurement patterns for different positions of the BD.

| ID | I1 | I2 | I0 | V0 | I4 | I5 |
|----|----|----|----|----|----|----|
| P1 | -  | -  | ~0 | ~0 | ~0 | ~0 |
| P2 | +  | -  | ~0 | ~0 | ~0 | ~0 |
| P3 | ~0 | +  | ~0 | ~0 | ~0 | ~0 |
| P4 | ~0 | +  | ~0 | ~0 | N  | N  |
| P5 | ~0 | +  | N  | N  | ~0 | ~0 |

On the basis of Table II, it should be possible to discriminate BDs at higher voltage, as all positions present different patterns. A BD near the vessel (SCD) comes always with noisy measurements I0 and V0, whereas the sign of the I1 and I2 first peak helps discriminating between the first 4 positions. Actually position 4 is differentiated from position 3 just because measurements in HVD1 (Core Snubber) are found to be quite noisy in this case. Table II could become a useful tool to address the search for the BD position in the future, but attention has to be paid to the signal conditioning, especially for the measurements from the rogowski coils, which have to be adequately filtered in order to get the correct sign of the first peak. Phase 1 results basically confirmed up to 150kV what has been found in Phase 0 for position P5; it was possible to acquire interesting images of the intentional BD with the SCD with the slow camera (C1, Basler camera), showing the lights corresponding to the arcs in the movable contacts, see Fig. 11. The frames are integrated over a period of about 800ms, therefore we have no information about the jitter of the contacts. However, tests carried out with long oscilloscope acquisition time shows that a jitter exists, especially at low voltage. Therefore the simulation of a BD with the SCD is slightly different with respect to a real breakdown from -1MV to ground.

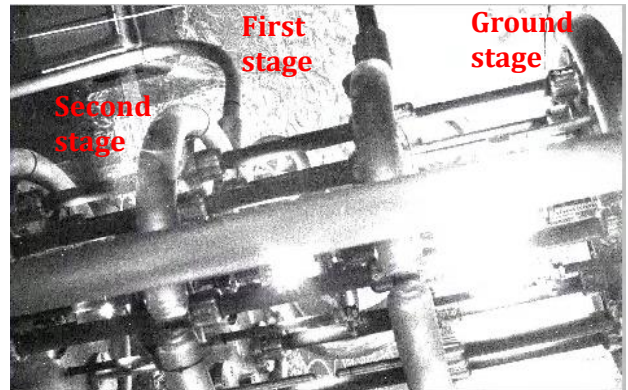

Fig. 11 – A frame of the slow camera C1 looking at the SCD, only some stages are shown. A BD between movable contacts is visible; the light is reflected by the internal wall of the vessel.

*A. Phase 2*

After the completion of Phase 1, the voltage has been increased to test the voltage withstanding capability of the system while being ready to capture any breakdown with the improved diagnostics. The first part of the test took place in April 2022 and has been performed with the SCD movable contacts permanently open, with a distance of the fixed spark-gaps contacts set to 50mm, a value which has been chosen to limit the overvoltage to a safe value for the HVB assembly. During this part of the test, the fast camera C0 was not yet installed. The TPS voltage has been increased in steps of about 100kV, each voltage level was applied for about 10 minutes before proceeding to the next level. During this process, the instrumentation in the HVH was moved around in predefined positions to search for discharge precursor in air [6]. Unfortunately, at about 850kV, a BD event was detected, without any precursor from partial discharge sensors in air or in $SF_6$. The TPS protection tripped, a sound was recorded by the microphones in the HVH and the slow camera showed an intense light coming from inside the vessel, which saturated the image. Instrumentation in HVH did not reveal any BD in air, nor corona effects which could anticipate a discharge. In this case, a BD has certainly occurred in vessel in proximity of the SCD, as the images from C1 clearly indicate. However, it was not clear why a sound was heard in the HVH; the sound was found to come out from the cavity between the HVBA and the HVD1 by the directional microphone installed in HVH [6] at about 20m distance. Concerning the measurements, I0 (current in the connection from the SCD to the vessel), shows a noisy pattern at the beginning and then a relatively slow transient about 35μs, with current as high as 200A; rogowski I2 (HVD1 bushing common mode current) has a positive first peak. Both features seem to suggest a scenario compatible with a BD in vessel from SCD to ground. I0 and I2 waveforms are reported in Fig. 12. The tests stopped after this event due to a fault in the TPS, as confirmed later on by the inspection. The inspection of the TL, HVD1 and SCD revealed no evident sign of discharge neither in air nor $SF_6$. Some marks and rust were found at the interface box between the TL2 and HVD1, but an analysis showed that they most likely were present before the event and could not be clearly associated with a BD in that position.

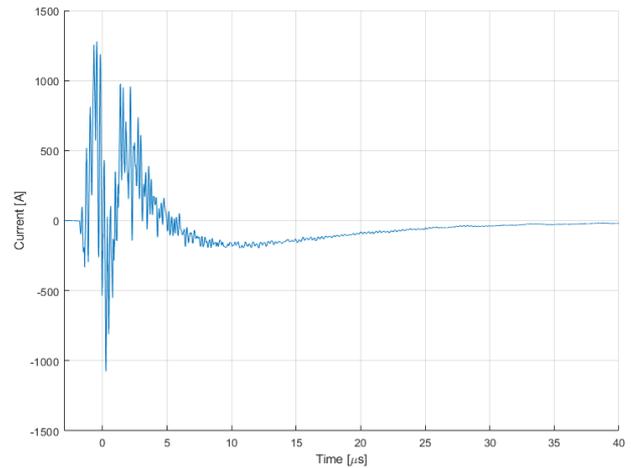

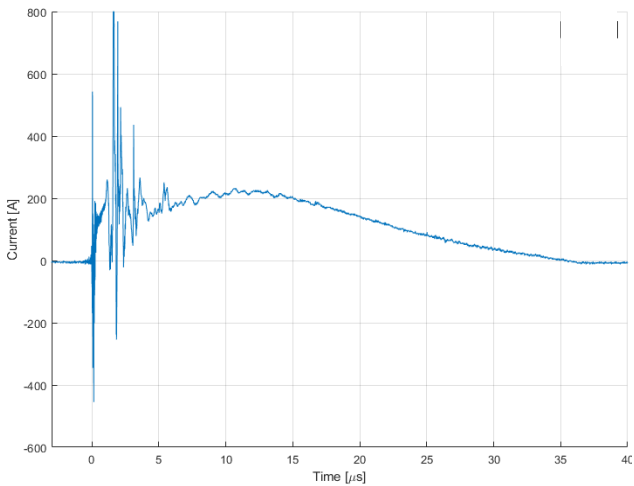

Fig. 12 – I0 (upper plot) and I2 (lower plot) signals recorded during the first event in Phase 2 tests. Both signals are compatible with a BD in vessel.

Inspection of the TPS showed signs of discharge in the output limiting resistor, likely due to the sudden overvoltage applied to this element at the onset of the BD.

The next round of tests started after the repair of the TPS, in July 2022. The diagnostic set was further enhanced with the addition of a fast camera (C0) looking at the vessel, with the introduction of an array of acoustic sensors all along the plant and the addition of more cameras in the HVH and inside the HVD1, pointing at the source of sound and other parts of the plant. After an analysis on the safe voltage limit, the length of the gaps has been increased to 70 mm to enhance the theoretical voltage withstanding capability of the equipment while preserving the protective function. The test was repeated as usual increasing the TPS voltage in steps and moving around the HVH partial discharge and corona sensors in search of a weak insulation area in air. Again, a BD was triggered at about 850kV, with the same pattern of events: light detected in the vessel, sound coming from the acoustic sensors, TPS protection tripping. No evidence of phenomena in air came from the HVH instrumentation also this time [6]. The fast camera was set to look at the fifth stage (1000kV-800kV) with a time resolution of about 1 μs, therefore we were able to capture the evolution of the BD in vessel, even though only a partial view was available. Fig. 13 shows a sequence of frames of the discharge, starting from the very first frame with light.

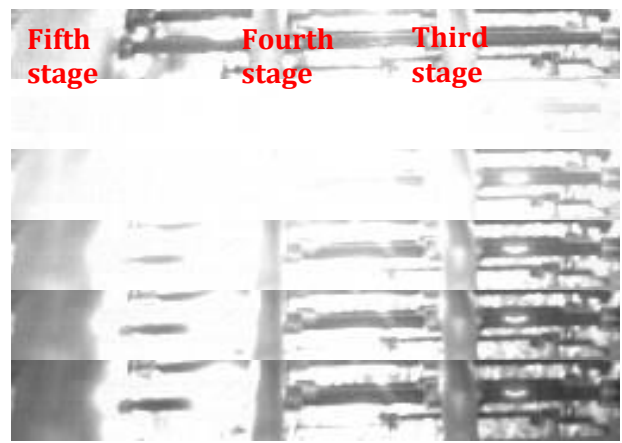

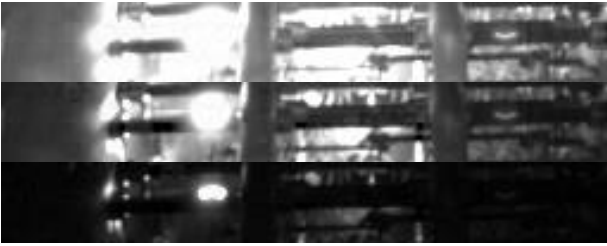

Fig. 13 – Sequence of pictures of the BD event occurred in July 2022, taken from the fast camera. Each picture shows a frame starting from the first frame with light.

In the first frame, it appears that the BD begins somewhere outside the framing position (maybe the fourth stage), and then a big light erupts from the fixed spark gap of the fifth stage. Following frames show that that the arc focuses just between the contacts of the fixed spark gap. No light is coming from other positions and finally the arc extinguishes in a relatively long time (tens of µs). The slow camera C1 was looking at the second stage (400-200kV) and below, the frame acquired is shown in Fig. 14. It just shows that the light of the arc comes from stages above and that no arc is seen for the second and first stage.

Concerning the sounds, unfortunately the data collected during the event were not useful as the microphones recorded a signal in many different positions, including MITICA bunker, HVH and external area nearby the TPS, apparently originated at the same time, while delay of some milliseconds was expected due to propagation. The reliability of these signals is doubtful as the microphones were not specifically designed for this application and electric noise could have had an impact on the acquisition.

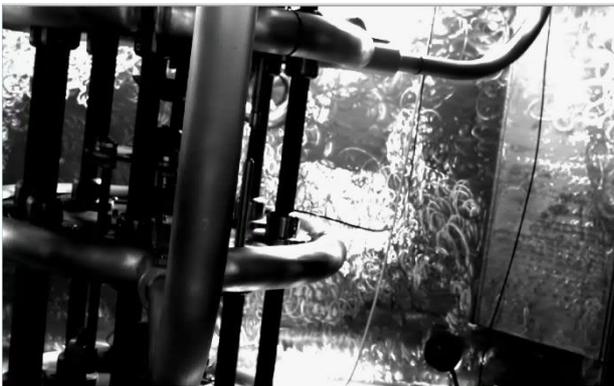

Fig. 14 – A frame of the slow camera C1 looking at the stage 400kV-200kV. The light seems to come from the region above the framing.

The measurement apparently showed a scenario compatible with a BD in vessel. The current I0, measured in the connection between the last stage of the SCD and the vessel, is shown in Fig. 15, compared to the same measurement obtained in the previous event. Note that the measurement has been improved and the large oscillations found in the previous case are not present anymore. However, a noise is anyway present at the beginning, as it is the case for a BD in vessel.

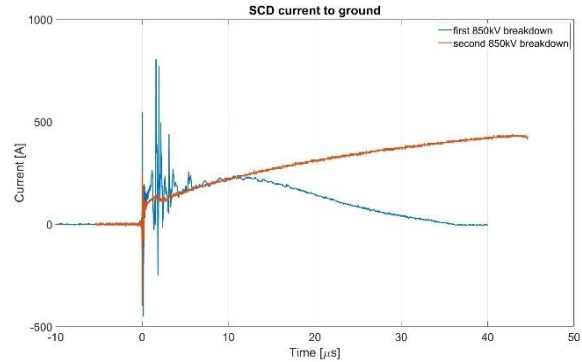

Fig. 15 – I0 signals recorded during the first event (blue curve) and second event (red curve).

The current seems to reach quite high values (up to 500A) and last for a relatively long time (probably more than 100 µs). The inspection following the event revealed marks compatible with a BD on fixed spark-gaps of the fifth (1000kV-800kV) and fourth (800kV-600kV) stages, but no other signs of BD elsewhere in $SF_6$ or air equipment. This is not compatible with the information coming from C0 and C1 cameras and with the current showed in Fig. 15, as it seems that there is no path for the circulation of this current. In fact, cameras and marks confirm that the arc is not present on all SCD fixed gaps and apparently there is no way for the current to reclose through the cable connection between the last SCD stage and the vessel, where I0 sensor is placed. This inconsistency between signals and camera information is so far unexplained. The analysis is still going on also considering all other measurements in order to find a self-consistent explanation and inspection on other parts previously not considered.

## 5. Discussion

The results of the tests with improved diagnostic system were very useful for the assessment of the BD location; however a fully coherent picture of the results is still missing, as some of them appear to be contradictory and have no explanation at the moment.

A solid evidence from the test is that the BD is not located in air, as the indications coming from the HVH instrumentation showed no sign of corona activity and no arcs in air were visible. Therefore the BD must be located in the $SF_6$ insulated equipment.

A BD has occurred in the vessel and in particular between the fixed spark gaps; this is confirmed by images, measurements and inspection. However, the following contradictory facts remain to be explained:

- the fixed spark gap length seems large enough to withstand at least 200kV; therefore BD between the contacts seems unlikely (a specific test has been carried out to confirm the withstanding capability of the first two stages);
- the sound located in HVH seems not compatible with a BD in vessel, unless a propagation is demonstrated;
- the current measured by I0 sensor on the connection between the last stage of the SCD and

the vessel does not correspond to a full BD of the SCD (the arc is found only on two stages); therefore the path of such current is unknown;

An interpretation of events has been suggested in the direction of addressing especially the first two points on the list. The assumption made by this interpretation is that the BD is not unique but actually multiple breakdowns (at least two) occurred in each event. A first BD would be triggered somewhere in $SF_6$, e.g. in the TL or in the HVBA, and then the overvoltage generated by the first BD would propagate to the SCD triggering a second BD between the fixed spark-gaps. Although this sequence would explain why the fixed spark-gaps triggered an arc and the sound heard in HVH, it fails to justify other facts, such that apparently, there is no mark anywhere in the $SF_6$ insulated equipment other than the SCD fixed spark gaps.

A clarification of the above contradiction requires more analysis and tests. At this stage, we are only able to discriminate the fact that the BD is for sure located not in the air insulated part.

## 6. Conclusions and future activities in MITICA

The integration tests of the AGPS demonstrated the ability of the system to generate and sustain a voltage up to 700kV up to 1000s, showing at the same time that a BD in the plant can give raise to overvoltage affecting different components. The experience so far in MITICA showed that, although the power supply system was designed following the relevant standards and taking adequate margins for the components, the complexity of the system in terms of technology and operation generated unforeseen scenarios not covered by any standard. The system is the first of its kind at such high level of voltage, power and complexity, therefore there is little experience in the fusion community for similar issues. The integration of the different components revealed the need for further investigation and protections: this was in fact part of the original MITICA mission. In this framework, the identification of the BD location is of paramount importance both in MITICA and in ITER: the existing set of diagnostic was not enough for this task and has been therefore improved, developing at the same time a strategy for BD identification which can also be applied in ITER.

The MITICA power supply plant has been improved enhanced diagnostic systems. The tests carried-out to identify the BD location have allowed to restrict the possible positions where the BD could happen. However, a clear and common understanding of the events is not reached; as data collected seems somehow in contradiction or give rise to unverified interpretations. The tests demonstrated that the BD is not located in the air insulated part; therefore the location of the BD must necessarily be in the $SF_6$ insulated parts. A BD happens in the SCD region, but whether this BD is unique or is the result of another BD happening somewhere else in $SF_6$ remains to be assessed.

Further analysis and tests are necessary. About the analyses, circuit models are under preparation to help the interpretation of the experimental results. Moreover, the electric field distribution inside the vessel in presence of the SCD equipped with the fixed spark-gaps is under review to assess the voltage withstanding capability of the entire assembly and find any possible weak point. Concerning the tests, there is room for improvement for enhancing the monitoring of the $SF_6$ insulation. The idea is to equip the plant with sensors able to identify partial discharge activities in SF6 with non-invasive methodology. Investigations are ongoing with prospective suppliers.

**Acknowledgement and Disclaimer**

This work has been carried out within the framework of the ITER-RFX Neutral Beam Testing Facility (NBTF) Agreement and has received funding from the ITER Organization. The views and opinions expressed herein do not necessarily reflect those of the ITER Organization.

This work has been carried out within the framework of the EUROfusion Consortium, funded by the European Union via the Euratom Research and Training Programme (Grant Agreement No 101052200 — EUROfusion). Views and opinions expressed are however those of the author(s) only and do not necessarily reflect those of the European Union or the European Commission. Neither the European Union nor the European Commission can be held responsible for them.